\pdfoutput=1

\documentclass[11pt]{article}

\usepackage[preprint]{acl}

\usepackage{times}
\usepackage{latexsym}

\usepackage[T1]{fontenc}

\usepackage[utf8]{inputenc}

\usepackage{microtype}

\usepackage{inconsolata}

\usepackage{amsmath}
\usepackage{courier}
\usepackage{bm}
\usepackage{algpseudocode}
\usepackage[noend,ruled,linesnumbered]{algorithm2e}
\usepackage{color}
\usepackage{enumitem}
\usepackage{url}
\usepackage{graphicx}
\usepackage{array}
\usepackage{multirow}
\usepackage{pifont}
\usepackage{xcolor,colortbl}
\usepackage{setspace}
\usepackage{makecell}
\usepackage{balance}
\usepackage{booktabs} 
\usepackage{amsthm}
\usepackage{amsfonts}
\usepackage{subcaption}
\usepackage{mdframed}
\usepackage{graphicx}
\usepackage{subcaption}

\newcommand{\ours}{\textsc{LogiCoL}\xspace}
\newcommand{\quest}{\textsc{Quest}\xspace}
\newcommand{\questvar}{\textsc{Quest+Varaiants}\xspace}

\definecolor{myblue}{rgb}{0.2, 0.2, 0.9}
\definecolor{myorange}{rgb}{0.9, 0.5, 0.0}

\usepackage{cleveref}
\crefformat{section}{\textcolor{blue}{\S#2#1#3}} 
\crefformat{subsection}{\textcolor{blue}{\S#2#1#3}}
\crefformat{subsubsection}{\textcolor{blue}{\S#2#1#3}}

\newcommand\blfootnote[1]{%
  \begingroup
  \renewcommand\thefootnote{}\footnote{#1}%
  \addtocounter{footnote}{-1}%
  \endgroup
}

\title{Set-Based Retrieval with Text Embedding Models}

\title{\ours: Logically-Informed Contrastive Learning \\ for Set-based Dense Retrieval}

\author{
Yanzhen Shen\textsuperscript{1*},
Sihao Chen\textsuperscript{2*},
Xueqiang Xu\textsuperscript{3},\\
\textbf{Yunyi Zhang}\textsuperscript{3},
\textbf{Chaitanya Malaviya\textsuperscript{4},
Dan Roth\textsuperscript{4}} \\
\textsuperscript{1}Stanford University, 
\textsuperscript{2}Microsoft \\ 
\textsuperscript{3}University of Illinois Urbana-Champaign,
\textsuperscript{4}University of Pennsylvania \\
\texttt{yanzhen4@stanford.edu}
}

\begin{document}

\maketitle
\blfootnote{*Work was done while the first two authors were affiliated with the University of Pennsylvania.}

\begin{abstract}
While significant progress has been made with dual- and bi-encoder dense retrievers, they often struggle on queries with logical connectives, a use case that is often overlooked yet important in downstream applications. Current dense retrievers struggle with such queries, such that the retrieved results do not respect the logical constraints implied in the queries.
To address this challenge, we introduce \ours, a logically-informed contrastive learning objective for dense retrievers. \ours builds upon in-batch supervised contrastive learning, and learns dense retrievers to respect the subset and mutually-exclusive set relation between query results via two sets of soft constraints expressed via t-norm in the learning objective. We evaluate the effectiveness of \ours on the task of entity retrieval, where the model is expected to retrieve a set of entities in Wikipedia that satisfy the implicit logical constraints in the query. We show that models trained with \ours yield improvement both in terms of retrieval performance and logical consistency in the results. We provide detailed analysis and insights to uncover why queries with logical connectives are challenging for dense retrievers and why \ours is most effective. Our codes and data are available at \href{https://github.com/yanzhen4/LogiCoL}{https://github.com/yanzhen4/LogiCoL}.

\end{abstract}
\section{Introduction}
\label{sec: Introduction}

In recent years, dense retrievers have become a prevalent class of methods for information retrieval (IR), leveraging learned text embeddings to compute the similarities between queries and documents \cite{karpukhin2020dense, izacard2021unsupervised}. Dense retrievers have led to significant advances across knowledge-intensive natural language processing (NLP) applications \cite{chen-etal-2017-reading, kwiatkowski-etal-2019-natural, petroni-etal-2021-kilt}.

\begin{figure}[t]
  \centering
  \includegraphics[width=\columnwidth]{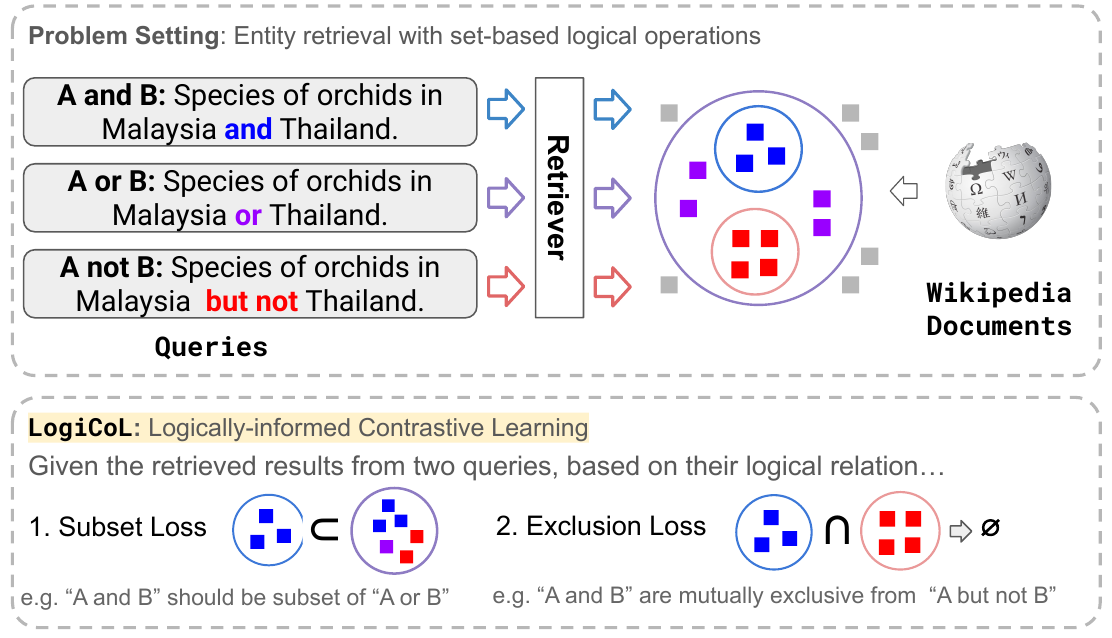}

  \caption{Overview of \ours. We study retrieval tasks where the queries involve set operations. With queries that form subset or mutually exclusive relation with each other, we formulate the logical consistency between the result sets as loss terms, on top of the in-batch contrastive learning objective of dense retrievers. }
  \vspace{-1em}
  \label{fig:overview}
\end{figure}

However, despite the progress made, dense retrievers today still struggle on user queries with logical structure \cite{malaviya2023quest}. 
In real-world applications, users often have complex retrieval needs involving multiple preferences or constraints.
As illustrated in \autoref{fig:overview}, user queries could contain different set operations, e.g. ``\textit{Species of orchids in Malaysia and/or/not in Thailand}''. 
While current dense retrievers are good at modeling the semantic relation between queries and documents, the logical relations within the query are often not well-represented in the embedding space, leading to logically irrelevant or even contradictory retrieval results \citep{malaviya2023quest, weller2024nevir, zhang2024boolquestions}. 

To address this challenge, we propose \ours, a \textbf{logi}cally-informed \textbf{co}ntrastive \textbf{l}earning framework that can be integrated with any dual- or bi-encoder dense retriever to enhance performance on queries with logical connectives. Our method builds upon in-batch supervised contrastive learning \cite{khosla2020supervised}, where we sample queries that share atomic sub-queries but with different logical connectives in the same mini-batch during training. For queries that we expect the retrieval results to be either subsets or mutually exclusive of each other, we train the model to respect the logical consistency between the result sets. The logical consistency constraints are expressed as two regularization terms based on t-norm \cite{li2019augmenting} in the learning objective, which model the subset and mutually exclusive relations between the retrieval results of queries.

We evaluate the effectiveness of \ours on the \quest dataset \cite{malaviya2023quest}, where given a query, a model is expected to rank and retrieve the set of entities in Wikipedia that satisfy the requirements and constraints in the query. 
Our experimental results show that \ours yields consistent improvement over off-the-shelf retrievers and supervised contrastive learning baselines in terms of retrieval performance and logical consistency of the retrieval results.   
By analyzing performance across different logical templates, we find that \ours brings the most significant improvements on complex queries involving implicit intersection and negation operators. This is particularly important because prior embedding models already perform reasonably well on queries involving union, where relevance is loosely defined, but struggle with intersection and especially negation. Compared to contrastive learning without logical constraints, \ours learns to separate queries that have different logical connectives but share a subset of ground truth documents, preventing logically distinct queries from collapsing to the same representation during model training. 

To summarize, our contributions are three-fold.
\begin{itemize}[leftmargin=*, nosep]
    \item We propose \ours, a logically-informed contrastive learning framework for retrieval with queries containing logical connectives. 
    \item We conduct experiments on the task of set-based entity retrieval, where the model is expected to retrieve a set of entities in Wikipedia that satisfy the implicit logical constraints in the query.   
    \item We provide detailed analysis and insights to uncover why queries with logical connectives are challenging for dense retrievers in general. 
\end{itemize}


\section{Motivation and Problem Setting}
\label{Motivation}
In this paper, we study retrieval tasks where the query contains logical connectives such as intersection (\textit{AND}), union (\textit{OR}), and negation (\textit{NOT}). Specifically, we focus on the task of entity retrieval, where a model is expected to retrieve a set of entities in Wikipedia that satisfy the implicit logical constraints in the query \cite{malaviya2023quest}.

\begin{figure}[t]
  \centering
  \includegraphics[width=0.90\columnwidth]{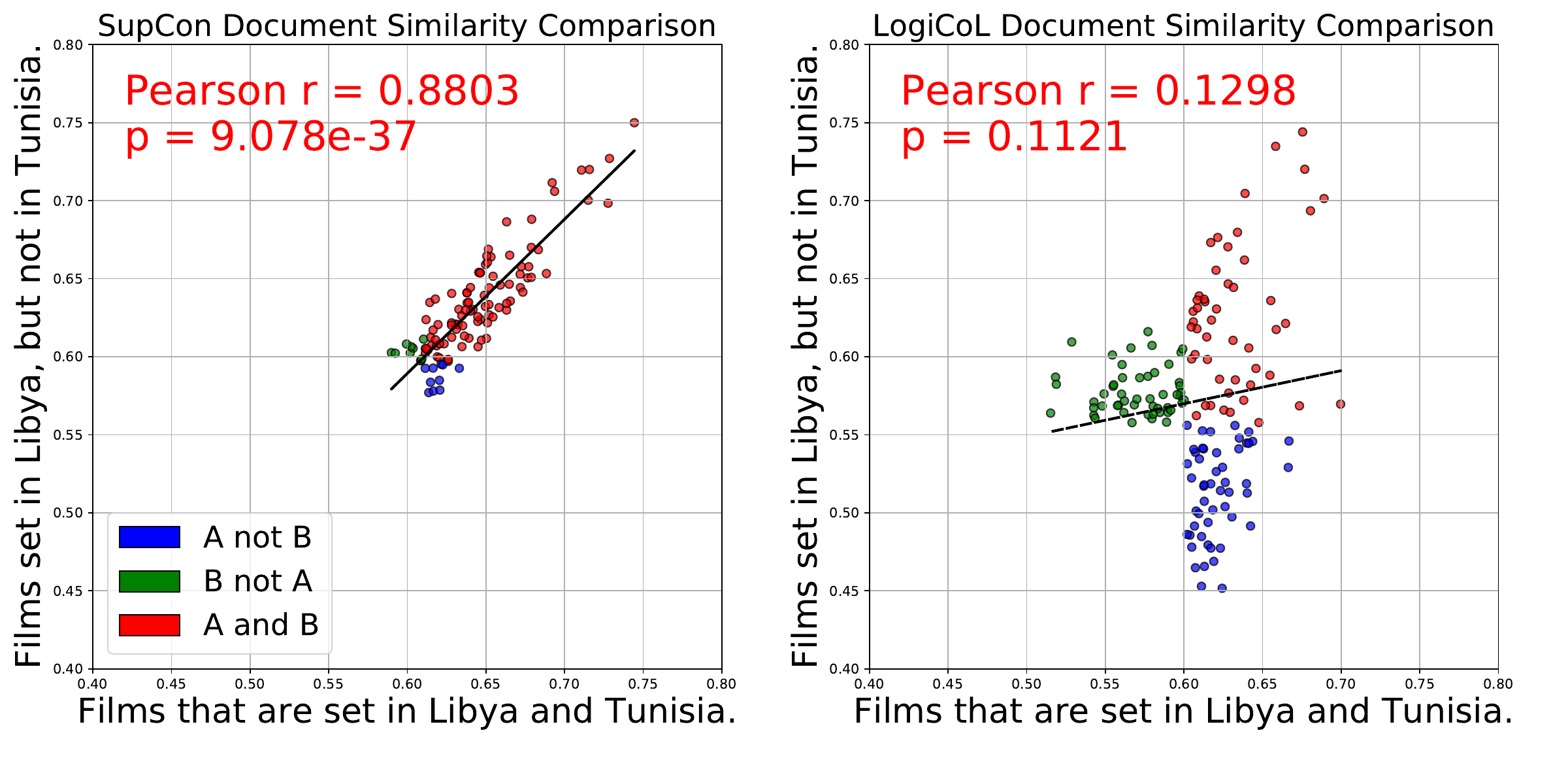}
  \vspace{-0.5em}
  \hspace{-2.1em}
  \caption{Document similarity correlation between queries of template \textit{A AND B} vs. \textit{A NOT B} of baseline (Left) and \ours (Right).}
  \label{fig:combined1}
  \vspace{-0.5em}
\end{figure}

\begin{figure}[t]
  \centering
  \includegraphics[width=0.97\columnwidth]{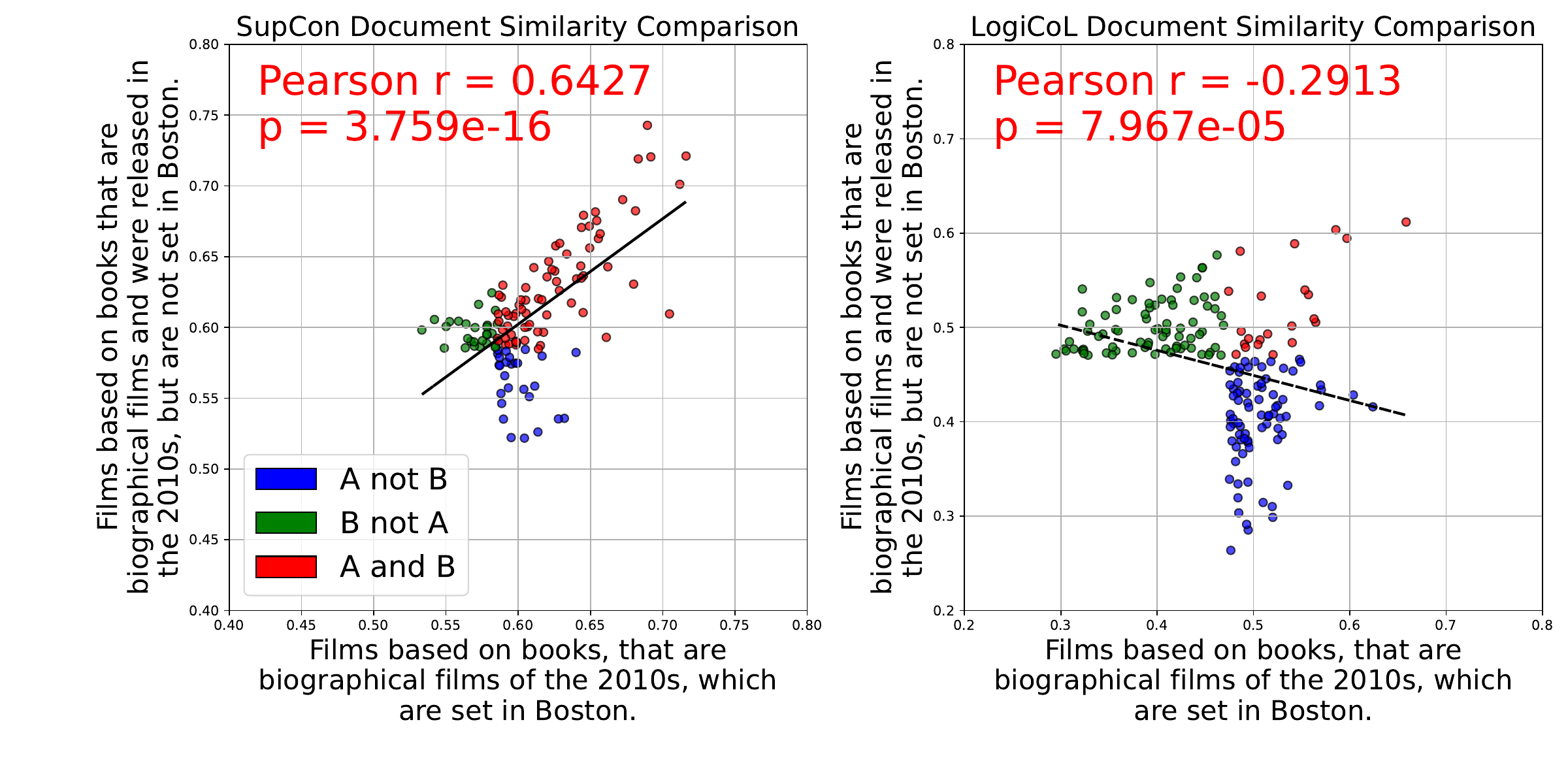}
  \vspace{-1em}
  \caption{Document similarity correlation between queries of template \textit{A AND B AND C} vs. \textit{A AND B NOT C} of baseline (Left) and \ours (Right). }
  \label{fig:combined2}
  \vspace{-0.5em}
\end{figure}

With current dense retriever models, we observe that their query embeddings cannot distinguish and capture such logical relations well.  To illustrate this, we visualize and compare the distributions of similarity scores between queries with different logical connectives and their top-100 retrieved documents. \autoref{fig:combined1} shows the similarity distribution for a query with template \textit{A AND B} (e.g., \textit{Films that are set in Libya and Tunisia}) versus \textit{A NOT B} (e.g., \textit{Films set in Libya, but not in Tunisia}). Similarly, in \autoref{fig:combined2}, we compare queries of template \textit{A AND B AND C} versus template \textit{A AND B NOT C}. In both examples, the subqueries A, B, and C are identical across the compared query pairs; the only difference lies in the logical structure. 

We observe that dense retrievers trained with the usual in-batch contrastive learning objectives cannot distinguish queries with different logical connectives. As shown in \autoref{fig:combined1} and \ref{fig:combined2}, the similarity score distribution for queries with different logical connectives is highly correlated ($r=.88$ and $.64$, respectively). This indicates that the query embeddings for different queries are highly similar, despite the fact that queries results should be mutually-exclusive.   
Moreover, there is substantial overlap in the top-100 retrieved documents, even though the expected results are disjoint sets. This pattern holds across both examples, highlighting the inability of current embedding models to encode logical distinctions. In comparison, models trained with \ours (\cref{sec:model}) show cleaner separation in the query embedding representations.

\section{\ours}
\label{sec:model}

In this section, we first describe the training data collection process (\cref{ssec:train-data}), and then the components of the \ours learning objective (\cref{ssec:contrastive}, \cref{ssec:constraints}).

\subsection{Sampling Logically Related Queries}
\label{ssec:train-data}
Our \ours approach builds on top of in-batch supervised contrastive learning \cite{khosla2020supervised}, and models the logical consistency between the retrieval results from queries with subset or exclusion relations. Compared to the common practice of randomly sampling queries into mini-batches for contrastive learning \cite{karpukhin2020dense, ni2021large}, \ours requires each mini-batch to contain queries with shared atomic sub-queries but different logical connectives. For example, if a mini-batch contains the query ``A and B'', then we would want ``A or B'' to be in the same mini-batch as well. 

We start from the \quest dataset, which features queries with annotations of logical connectives and atomic sub-queries. However, only a small portion ($\tilde~4\%$) of the queries in the dataset actually share atomic sub-queries with others. For this reason, we create an augmented version of \quest, where variants of queries with different logical connectives are included. We denote this augmented dataset as \questvar. The types of logical connectives used in \quest and \questvar are shown in \autoref{tab:data_statistics}. The ground truth document set for queries in \questvar is inferred from the set of ground truth entities for the atomic sub-queries.  More details on the construction of \questvar can be found in Appendix~\ref{app:questvar-details}.

\subsection{Contrastive Learning}
\label{ssec:contrastive}
During \ours training, each mini-batch is constructed by sampling queries sharing the same set of atomic sub-queries. For each query, we randomly sample one ground truth document as part of the mini-batch. As different queries could share some ground truth documents, each query could have more than one ground truth document in the mini-batch, which the commonly adopted InfoNCE loss \cite{oord2018representation} does not support. For such reason, we leverage and modify the supervised contrastive loss \cite{khosla2020supervised} as our base learning objective for \ours.

\begin{equation*}\small
\mathcal{L} = \sum_{i \in \mathcal{I}_q} \frac{-1}{|P(i)|} \sum_{p \in P(i)} \log \frac{\exp(\mathbf{q}_i \cdot \mathbf{d}_p / \tau)}{\sum_{j \in \mathcal{I} \setminus \{i\}} \exp(\mathbf{q}_i \cdot \mathbf{d}_j / \tau)}
\end{equation*}

Here, $P(i)$ denotes the set of indices of all positive documents to the $i$-th query within the mini-batch, and $|P(i)|$ is its cardinality. $\mathcal{I} = \{1, \ldots, N\}$ denote the set of all queries and document indices. $\mathbf{q}_i \in \mathbb{R}^d$ denote the encoded representation of the $i$-th query in the mini-batch, and let $\mathbf{d}_i \in \mathbb{R}^d$ denote the encoded representation of the $i$-th document in the mini-batch.

\subsection{Joint Constrained Learning}
\label{ssec:constraints}
On top of the base contrastive learning objective, \ours uses two regularization objectives to encourage the model to retrieve results that are consistent with the logical relations between queries. We specify two types of consistency requirements: exclusion relation consistency and subset relation consistency. 

\subsubsection{Exclusion Consistency}

Given any two queries \(q_1\) and \(q_2\) within a mini-batch, if they exhibit an exclusion relation, then their retrieval results should be disjoint. Intuitively, the query such as \textit{Orchids of Malaysia but not Thailand} and the query such as \textit{Orchids of Thailand} should not retrieve overlapping documents. 

To operationalize this constraint during training, we model the retrieval behavior of queries through their similarity distributions over documents, and require that any two queries with an exclusion relation should have divergent similarity distributions. Let \( q_i \) and \( q_j \) be two queries with an exclusion relation, and let \( \mathbf{q}_i \), \( \mathbf{q}_j \), and \( \mathbf{d}_k \) denote their encoded representations and that of a document \( d_k \) in the same training mini-batch.

We compute the cosine similarity scores \( s_{q_i,d_k} \) and \( s_{q_j,d_k} \) between each query and document \( d_k \) and normalize them using a softmax to obtain probability distributions \( \mathbf{s}_{q_i} \) and \( \mathbf{s}_{q_j} \) over the mini-batch of documents:
\[
\mathbf{s}_{q} = \text{softmax}([s_{q,d_1}, s_{q,d_2}, \dots, s_{q,d_N}]).
\]
To enforce divergence between the distributions of exclusion-related queries, we compute the symmetric Kullback--Leibler (KL) divergence between the two distributions:
\begin{equation}\small
\text{SymKL}(\mathbf{s}_{q_i}, \mathbf{s}_{q_j}) = \frac{1}{2} \left( \text{KL}(\mathbf{s}_{q_i} \,\|\, \mathbf{s}_{q_j}) + \text{KL}(\mathbf{s}_{q_j} \,\|\, \mathbf{s}_{q_i}) \right)
\end{equation}
where $\text{SymKL}(\cdot, \cdot)$ denotes the symmetric KL divergence, and $\text{KL}(p \,\|\, q)$ is the standard (asymmetric) KL divergence from distribution $p$ to $q$.

We then train the model using the following margin-based loss function:
\[
\mathcal{L}_{E}(q_i, q_j) = \max\left( \gamma_e - \text{SymKL}(\mathbf{s}_{q_i}, \mathbf{s}_{q_j}), 0 \right),
\]
where $\gamma_e > 0$ is a margin hyperparameter. The final exclusion loss averages over all exclusion-related query pairs within a mini-batch.

By maximizing the symmetric KL divergence between exclusion-related queries, we encourage the model to produce non-overlapping retrieval distributions, thereby enforcing exclusion consistency during training.

\subsubsection{Subset Consistency}

We define a subset relation between two queries \( (q_1, q_2) \) based on their logical structure. Specifically, \( q_1 \subset q_2 \) if the logical expression of \( q_1 \) implies that of \( q_2 \). For example, if \( q_1 = \textit{A AND B} \) and \( q_2 = \textit{A} \), then any document relevant to \( q_1 \) must also be relevant to \( q_2 \), since satisfying both \textit{\( A \)} and \textit{\( B \)} necessarily satisfies \( A \). The subset relation between two queries can be interpreted as a logical implication: for any document \(d\) in the mini-batch, if \(d\) is relevant to \(q_1\), then it should also be relevant to \(q_2\). This implies that a high similarity score between \(q_1\) and \(d\) must entail a correspondingly high similarity between \(q_2\) and \(d\). On the other hand, a case where \(\text{sim}(q_1, d)\) is high but \(\text{sim}(q_2, d)\) is low indicates a violation of this logical constraint. 

To rewrite this constraint in a differentiable loss function, we use the product t-norm and transformation to the negative log space as in previous works \cite{li2019logic, li2019augmenting}, we define the subset loss:

{\small
\begin{equation}
\begin{split}
\mathcal{L}_S = \sum_{(q_1, q_2) \in \mathcal{E}_S} \sum_{d \in \mathcal{D}} 
\max\Big( & \log \text{sim}(q_1, d) 
- \log \text{sim}(q_2, d) \\
& + \gamma_s, \; 0 \Big),
\end{split}
\end{equation}
}
where \(\gamma_s\) is a margin hyperparameter. 

Through this subset loss, we enforce the subset logical consistency, that is, for any \(d\), it imposes a penalty when \(\text{sim}(q_1, d)\) is high but \(\text{sim}(q_2, d)\) remains comparatively low, thereby violating the implication that relevance to \(q_1\) should entail relevance to \(q_2\).

\subsubsection{Joint Learning Objective}

After expressing the logical consistency requirements through additional margin-based loss terms, we combine all objectives into the following joint learning objective:
\begin{equation}
\mathcal{L}_{\text{joint}} = \mathcal{L} + \lambda_E \mathcal{L}_E + \lambda_S \mathcal{L}_S,
\end{equation}
where \(\mathcal{L}\) denotes the supervised contrastive loss, \(\mathcal{L}_E\) denotes the exclusion consistency loss, and \(\mathcal{L}_S\) denotes the subset consistency loss. The \(\lambda\) coefficients are non-negative hyperparameters to control the relative influence of each loss term.

\begin{table}[t]
\centering
\label{tab:main_experiment}
\resizebox{\linewidth}{!}{
\begin{tabular}{lccccccc}
\toprule
\small
\multirow{2}{*}{}{\textbf{Templates}} & \multicolumn{3}{c}{\textbf{\quest}} & \multicolumn{3}{c}{\textbf{\questvar}} \\
                                 & \textbf{Train} & \textbf{Validation} & \textbf{Test} & \textbf{Train} & \textbf{Validation} & 
                                 \textbf{Test}\\
\midrule
A                    &3482	&56	&260	&3992	&558	&2286      \\
$A \cap B$ & 168 & 54 & 271 & 430 & 278 & 1497 \\
$A \cup B$ & 78 & 44 & 260 & 498 & 458 & 2532 \\
$A \setminus B$ & 70 & 44 & 212 & 914 & 778 & 4226 \\
$A \cap B \cap C$ & 190 & 40 & 236 & 114 & 52 & 294 \\
$A \cap B \setminus C$ & 68 & 44 & 236 & 96 & 58 & 353 \\
$A \cup B \cup C$ & 38 & 41 & 252 & 118 & 113 & 666 \\
Total & 4094 & 323 & 1727 & 6162 & 2295 & 11854 \\

\bottomrule
\end{tabular}
}
\vspace{-0.5em}
\caption{Types of queries and dataset statistics for \quest and \questvar.}
\vspace{-1em}
\label{tab:data_statistics}
\end{table}

\subsection{Mixture of Random and Related Query Batching}
To better understand the impact of changing the batch strategy compared to random batching for \ours, we introduce a mixed batching strategy, where a mini-batch is comprised of both randomly sampled queries and groups of related queries. We use a hyperparameter $\alpha$ to control the proportion of random samples in the mixture. An analysis of the effect of $\alpha$ is discussed in \cref{sec:grouping_randomness}.

\section{Experiments} 

\begin{table*}[!t]
\centering
\renewcommand{\arraystretch}{1.15}
\small
\resizebox{\textwidth}{!}{%
\begin{tabular}{c|c|ccccc|ccccc}
\toprule
\multirow{2}{*}{\textbf{Backbone}} & \multirow{2}{*}{\textbf{Method}} 
& \multicolumn{5}{c|}{\textbf{\quest}} 
& \multicolumn{5}{c}{\textbf{\questvar}} \\
\cmidrule(lr){3-7} \cmidrule(lr){8-12}
& & P@1 & R@5 & R@20 & R@100 & R@1000
  & P@1 & R@5 & R@20 & R@100 & R@1000 \\
\midrule
\multicolumn{2}{c|}{\textbf{BM25}} & 11.64 & 4.73 & 10.41 & 19.68 & 39.49 & 6.62 & 1.92 & 4.40 & 9.32 & 22.51 \\
\midrule
\multirow{3}{*}{\textbf{GTR-base}}
& Zero-shot & 8.98 & 3.41 & 7.07 & 13.46 & 30.95 & 5.63 & 1.38 & 3.13 & 6.66 & 18.82 \\
& SupCon & 19.75 & 8.08 & 16.66 & 31.71 & 60.34 & 12.38 & 3.45 & 7.45 & 15.63 & 38.04 \\
& \ours & \textbf{20.21} & \textbf{8.36} & \textbf{17.23} & \textbf{33.08} & \textbf{61.86} & \textbf{13.84} & \textbf{3.62} & \textbf{8.07} & \textbf{17.22} & \textbf{41.93} \\
\midrule
\multirow{3}{*}{\textbf{GTE-base}}
& Zero-shot & 19.40 & 7.79 & 16.18 & 31.81 & 61.87 & 11.28 & 3.18 & 7.11 & 15.75 & 38.97 \\
& SupCon & \textbf{25.77} & \textbf{11.16} & 22.50 & 39.58 & 69.66 & 14.61 & 4.18 & 9.61 & 19.60 & 44.82 \\
& \ours & 25.65 & 10.98 & \textbf{22.60} & \textbf{40.69} & \textbf{71.24} & \textbf{15.78} & \textbf{4.50} & \textbf{10.14} & \textbf{21.46} & \textbf{48.87} \\
\midrule
\multirow{3}{*}{\textbf{Contriever}}
& Zero-shot & 7.70 & 3.01 & 6.37 & 13.64 & 32.78 & 5.61 & 1.39 & 3.18 & 7.33 & 20.96 \\
& SupCon & \textbf{21.31} & 9.59 & 20.41 & 38.00 & 69.30 & 13.38 & 3.71 & 8.40 & 18.23 & 43.94 \\
& \ours & 20.56 & \textbf{9.76} & \textbf{20.45} & \textbf{39.38} & \textbf{71.46} & \textbf{13.84} & \textbf{4.03} & \textbf{9.50} & \textbf{20.71} & \textbf{49.16} \\
\midrule
\multirow{3}{*}{\textbf{E5-base-v2}}
& Zero-shot & 18.18 & 7.75 & 16.29 & 30.97 & 60.19 & 11.45 & 3.09 & 6.90 & 14.99 & 37.70 \\
& SupCon & \textbf{26.29} & 11.16 & 22.47 & 40.04 & 70.32 & 15.26 & 4.48 & 9.58 & 19.82 & 45.49 \\
& \ours & 24.32 & \textbf{11.70} & \textbf{23.48} & \textbf{42.13} & \textbf{73.52} & \textbf{16.27} & \textbf{4.67} & \textbf{10.38} & \textbf{21.87} & \textbf{50.24} \\
\bottomrule
\end{tabular}%
}
\vspace{-0.5em}
\caption{Performance comparison on \quest and \questvar datasets across different backbone models and methods. Best scores per backbone model are bolded.}
\vspace{-0.5em}
\label{tab:main_experiment}
\end{table*}

\subsection{Dataset}
We evaluate \ours using the \quest benchmark dataset \cite{malaviya2023quest}. \quest consists of 4,094 training queries and 1,727 testing queries spanning the domains of films, books, plants, and animals. The augmented \questvar contains 6,162 training queries and 11,854 testing queries. A detailed breakdown of the original and augmented Quest datasets across different templates is in Table \ref{tab:data_statistics}.

\subsection{Model Configurations}
We initialize the transformer encoder layers with pre-trained weights from four types of sentence encoders: GTR \cite{ni2021large}, Contriever \cite{izacard2021unsupervised}, GTE \cite{li2023towards}, and E5 \cite{wang2022text}. For each encoder, we use the base version of the model. In the case of E5, we specifically adopt its latest second version, E5-base-v2, to ensure stronger baseline performance.

To demonstrate the effectiveness of our framework, we use the original \quest training queries and documents to fine-tune the sentence encoders using contrastive loss with random batching, which we denote as \textit{SupCon}. 
We also include BM25 \cite{robertson2009probabilistic} for reference. 

\subsubsection{Evaluation Metrics}
Given an encoder's ranking of over 300k documents in the corpus for a given test query, we evaluate the results by Precision@1, denoted as P@1, and Recall@\{5, 20, 100, 1000\}, denoted as R@n, against the ground truth set of relevant documents.

\subsection{Implementation Details}
We train our models on four NVIDIA TITAN RTX GPUs for 10 epochs with an average effective batch size of 16. We adopt a constant learning rate for all four backbone models. For GTE \cite{li2023towards}, Contriever \cite{izacard2021unsupervised}, and E5 \cite{wang2022text}, we used the recommended learning rate used in the original paper, which are 2e-5, 1e-5, and 1e-5. For GTR \cite{ni2021large}, we set the learning rate to be 1e-5, which significantly outperforms the original recommended learning rate 1e-3 using both \textit{SupCon} and \ours frameworks. In our joint constraint learning, we tune its hyperparameters based on performance on the development set. $\lambda_E = 0.1$ and $\lambda_S = 0.1$, and we set $\gamma_e = 0.2$ and $\gamma_e = 0.2$.

\subsection{Experiment Results}
Table \ref{tab:main_experiment} presents the evaluation results on both \quest and \questvar. We observe that applying the \ours fine-tuning framework consistently enhances performance across different backbone encoders, outperforming both the zero-shot baselines and \textit{SupCon}. Notably, \ours yields the most significant gains on Recall@1000, improving by approximately 2\% to 3\% over the fine-tuned baselines on the original test set, and by around 5\% on the augmented test set. Given that the test sets contain a substantial number of queries (1,727 and 11,854, respectively), each associated with hundreds of ground-truth documents, these improvements provide strong evidence for the effectiveness of our framework.

We categorize queries into four types based on their implicit logical operators: None (i.e., single queries), Intersection, Negation, and Union. We then compare the performance of \ours against the baseline across these categories. As shown in Table \ref{tab:performance_across_templates}, \ours consistently outperforms the baselines on atomic queries as well as complex queries involving intersection and negation operators. Although there is a slight drop in union queries, this can be attributed to the fact that the dense embeddings learned from the baseline contrastive learning schema treat all logical operators as the union operator, which leads to marginally higher performance on the union queries and much lower performance on the intersection and negation queries in comparison to models trained using \ours. 

\subsection{Ablation Study}
The key technical novelty of \ours is twofold: (1) we use groups of related queries as the smallest unit in our contrastive learning batch, enabling the model to learn the logical connectives within queries and logical relations between queries. (2) We apply joint constrained learning to enforce logical consistency among in-batch queries. We demonstrate the contributions of these techniques through a comprehensive ablation study. Specifically, we examine the following ablated variants:
\begin{itemize}[leftmargin=*,nosep]
    \item \textbf{SupCon}: Trained on the original \quest data using random batching.
    \item \textbf{\ours\ -- GroupStrategy -- Constraints}: Trained on \questvar training data using random batching.
    \item \textbf{\ours\ -- MixStrategy -- Constraints}: Trained on \questvar using all grouped batches.
    \item \textbf{\ours\ -- Constraints}: Trained on \questvar using the mix strategy, but without applying joint constraint learning.
\end{itemize}

We compare these four ablated versions against \ours. All variants are evaluated using the two strongest backbone models, Contriever and E5-base-v2, and we report performance in terms of Recall@100 (R@100) and Recall@1000 (R@1000). Our results in Table \ref{tab:ablation} reveal the following key observations: (1) \textbf{\ours\ -- GroupStrategy -- Constraints} provides minimal improvement over \textbf{SupCon}, suggesting that simply increasing the number of training queries offers little benefit. (2)  \textbf{ \ours\ -- Constraints} improves performance over \textbf{\ours\ -- MixStrategy -- Constraints} when using the E5-base-v2 backbone, but underperforms it when experimenting with Contriever. This suggests that solely using batches of queries that share atomic sub-queries has the risk of learning insufficient semantic information. (3) \textbf{\ours\  - Constraints} consistently and significantly outperforms both \textbf{\ours\ -- GroupStrategy -- Constraints} and \textbf{SupCon}, underscoring the advantage of our mixed group-based batching strategy that leverages both logical and semantic similarities. (4) Finally, \ours further improves upon \textbf{\ours\ -- Constraints}, demonstrating that enforcing logical consistency through joint constrained learning provides consistent gains during training.

\begin{table}[t]
\centering
\renewcommand{\arraystretch}{1.15}
\resizebox{1\linewidth}{!}{
\small
\begin{tabular}{c|c|ccccc}
\toprule
\multirow{2}{*}{\textbf{Backbone}} & \multirow{2}{*}{\textbf{Method}} 
& \multicolumn{5}{c}{\textbf{Performance across Query Templates (\%)}} \\
\cmidrule(lr){3-7}
& & None & Intersection & Negation & Union & All \\
\midrule
\multirow{2}{*}{\textbf{GTR-base}} 
& SupCon & 22.10 & 18.02 & 14.71 & 10.68 & 15.63 \\
& \ours       & \textbf{23.61} & \textbf{19.34} & \textbf{17.10} & \textbf{11.35} & \textbf{17.22} \\
\midrule
\multirow{2}{*}{\textbf{GTE-base}}
& SupCon & 26.97 & 23.24 & 18.04 & \textbf{13.99} & 19.60 \\
& \ours       & \textbf{29.30} & \textbf{27.37} & \textbf{21.46} & 12.45 & \textbf{21.46} \\
\midrule
\multirow{2}{*}{\textbf{Contriever}}
& SupCon & 26.06 & 21.94 & 15.71 & \textbf{13.54} & 18.23 \\
& \ours       & \textbf{27.78} & \textbf{26.62} & \textbf{20.43} & 12.23 & \textbf{20.71} \\
\midrule
\multirow{2}{*}{\textbf{E5-base-v2}}
& SupCon & 26.91 & 22.85 & 18.92 & \textbf{13.93} & 19.82 \\
& \ours       & \textbf{29.01} & \textbf{27.20} & \textbf{21.48} & 13.86 & \textbf{21.87} \\
\bottomrule
\end{tabular}
}
\vspace{-0.5em}
\caption{Recall@100 across different backbone models on \questvar queries with different logical connectives. }
\vspace{-0.5em}
\label{tab:performance_across_templates}
\end{table}

\begin{table}[htbp]
\centering
\renewcommand{\arraystretch}{1.15}
\resizebox{1\linewidth}{!}{
\small
\begin{tabular}{l|cc|cc}
\toprule
\textbf{Method} & \multicolumn{2}{c|}{\textbf{Contriever}} & \multicolumn{2}{c}{\textbf{E5-base-v2}} \\
\cmidrule(lr){2-3} \cmidrule(lr){4-5}
& \textbf{R@100} & \textbf{R@1000} & \textbf{R@100} & \textbf{R@1000} \\
\midrule
 SupCon & 38.00 & 69.30 & 40.04 & 70.32 \\
\midrule
\hspace{1em} -- GroupStrategy -- Constraints & 38.34 & 70.41 & 39.87 & 69.94 \\
\hspace{1em} -- MixStrategy -- Constraints & 38.05 & 70.33 & 40.46 & 72.14 \\
\hspace{1em} -- Constraints & 39.21 & 71.39 & 41.73 & 73.26 \\
\midrule
\ours & \textbf{39.38} & \textbf{71.46} & \textbf{42.13} & \textbf{73.52} \\
\bottomrule
\end{tabular}
}
\vspace{-0.5em}
\caption{Ablation versions of \ours on \quest queries. Rows below show versions with specific components removed.}
  \vspace{-1em}
\label{tab:ablation}
\end{table}

\subsection{Document Similarity Distribution}

To further validate the effectiveness of our \ours framework, we repeat the same analysis from Section \cref{Motivation}, comparing document similarity distributions for logically mutually-exclusive queries under the same retrieval setting described in Section~3. In Figures ~\ref{fig:combined1} and ~\ref{fig:combined2}, it shows that \ours significantly reduces the similarity correlation between such queries in comparison to the baseline's result. The Pearson coefficients drop to $0.1298$ and even $-0.2913$, indicating that the model no longer treats logically mutually-exclusive queries as semantically equivalent. The retrieved documents are also more cleanly separated, with minimal overlap in the top-100 results. These findings demonstrate that \ours effectively encodes logical structure in the embedding space, enabling more faithful retrieval aligned with query intent.
\section{Analysis and Discussion}

\subsection{Evaluating Logical Consistency in Retrieved Results}

In this analysis, we aim to evaluate whether for queries with negation operators ($A \setminus B$ and $A \cap B \setminus C$) models trained using \ours can better avoid retrieving logically excluded documents ($B$ and $C$ respectively). To measure this, we compute the average ranking positions of ground-truth documents in the retrieval results $\text{AvgRank}^+_q$ and the average ranking positions of the logically excluded documents $\text{AvgRank}^-_q$. 
Then, if the logically excluded documents are ranked higher than its ground truth documents, which is $\text{AvgRank}^-_q < \text{AvgRank}^+_q$, then it is said that the retrieval result of this query has violated the logical consistency.

We define the overall violation rate across the test set as:

{\small
\begin{equation}
\text{ViolationRate} = \frac{1}{|\mathcal{Q}|} \sum_{q \in \mathcal{Q}} 
\mathbb{I} \left[ \text{AvgRank}^-_q < \text{AvgRank}^+_q \right]
\end{equation}
}

\begin{figure}[t]  
  \centering
  \includegraphics[width=\columnwidth]{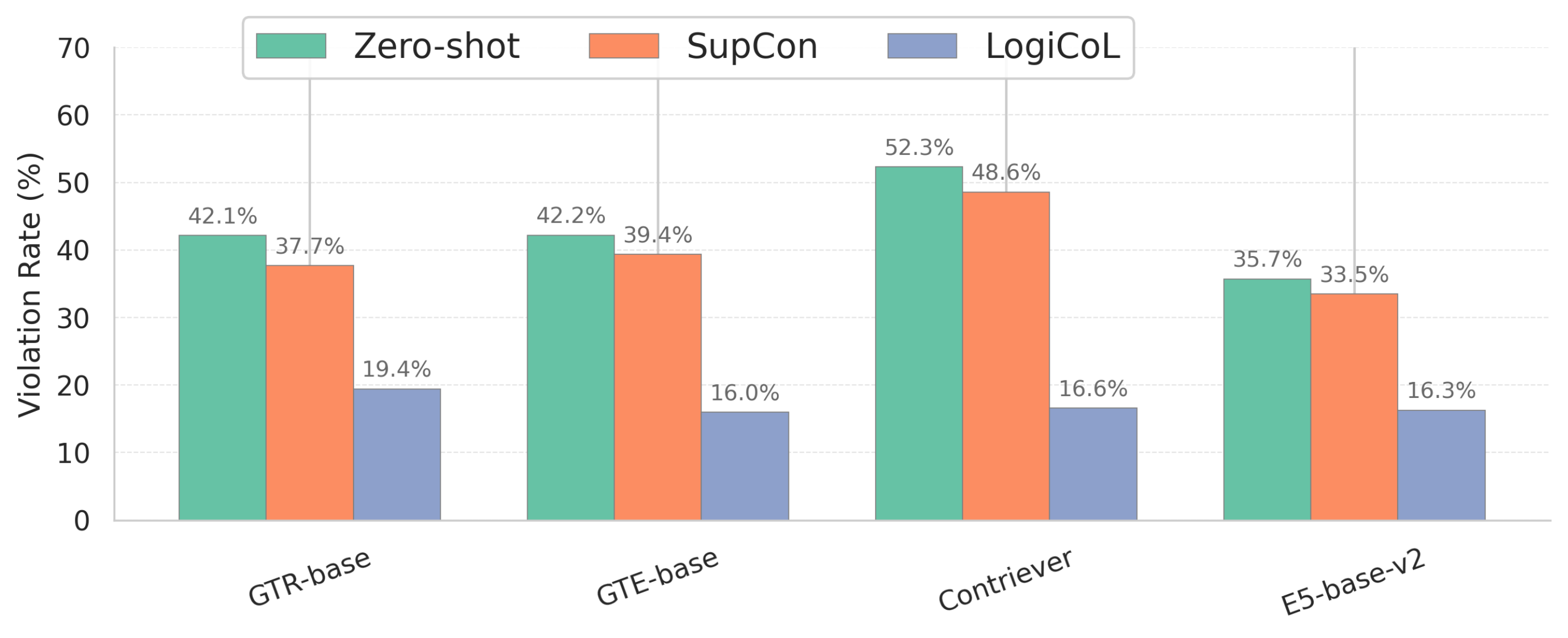}
\vspace{-0.5em}
  \caption{Logical consistency violation rate on queries with negation in comparison with \textit{Zero-shot} and \textit{SupCon} baselines on \questvar. Higher value indicates higher violation rate.}
  \label{fig: violation}
\end{figure}

We show our result in Figure \ref{fig: violation}. \textit{Zero-shot} and \textit{SupCon} both have high violation rates, implying that current models fail to identify the exclusion relations, and this skill cannot be learned through simple supervised contrastive learning. On the other hand, training with \ours framework significantly decreases the violation rate by over 20\% for all backbone models, indicating they have successfully learned this skill. 

\subsection{Effect of Grouping Randomness on Embedding Coherence and Performance}
\label{sec:grouping_randomness}

\begin{figure}[t]
  \centering
  \includegraphics[width=\columnwidth]{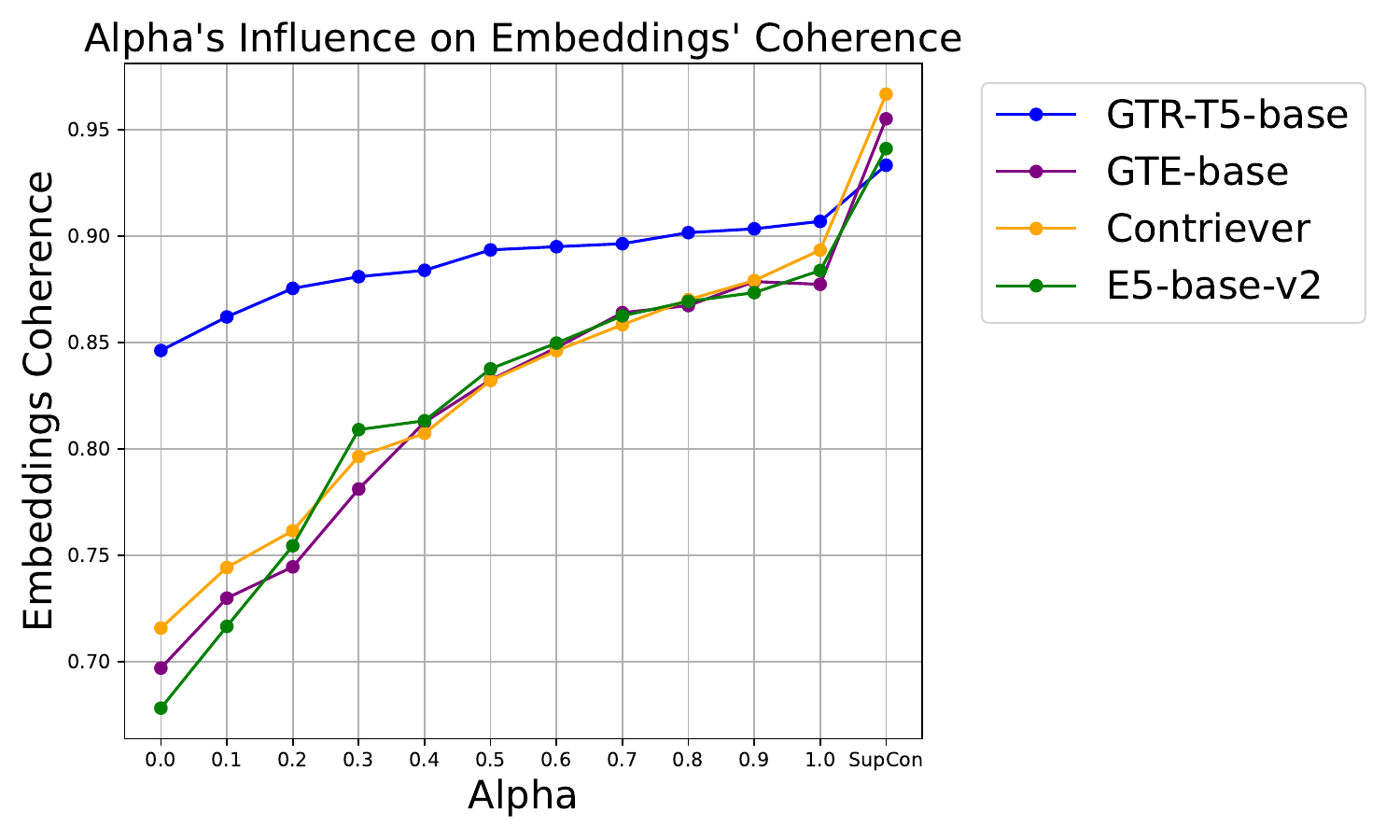}
\vspace{-0.5em}
  \caption{Coherence of embeddings of \ours with different $\alpha$ values, which indicates the randomness in batching strategy on \questvar. Coherence of \textit{SupCon} baseline is put after the result of $\alpha$ = 1 for reference. }
  \label{fig:hyperparam_study}
\end{figure}

\begin{figure}[t]
  \centering
  \includegraphics[width=\columnwidth]{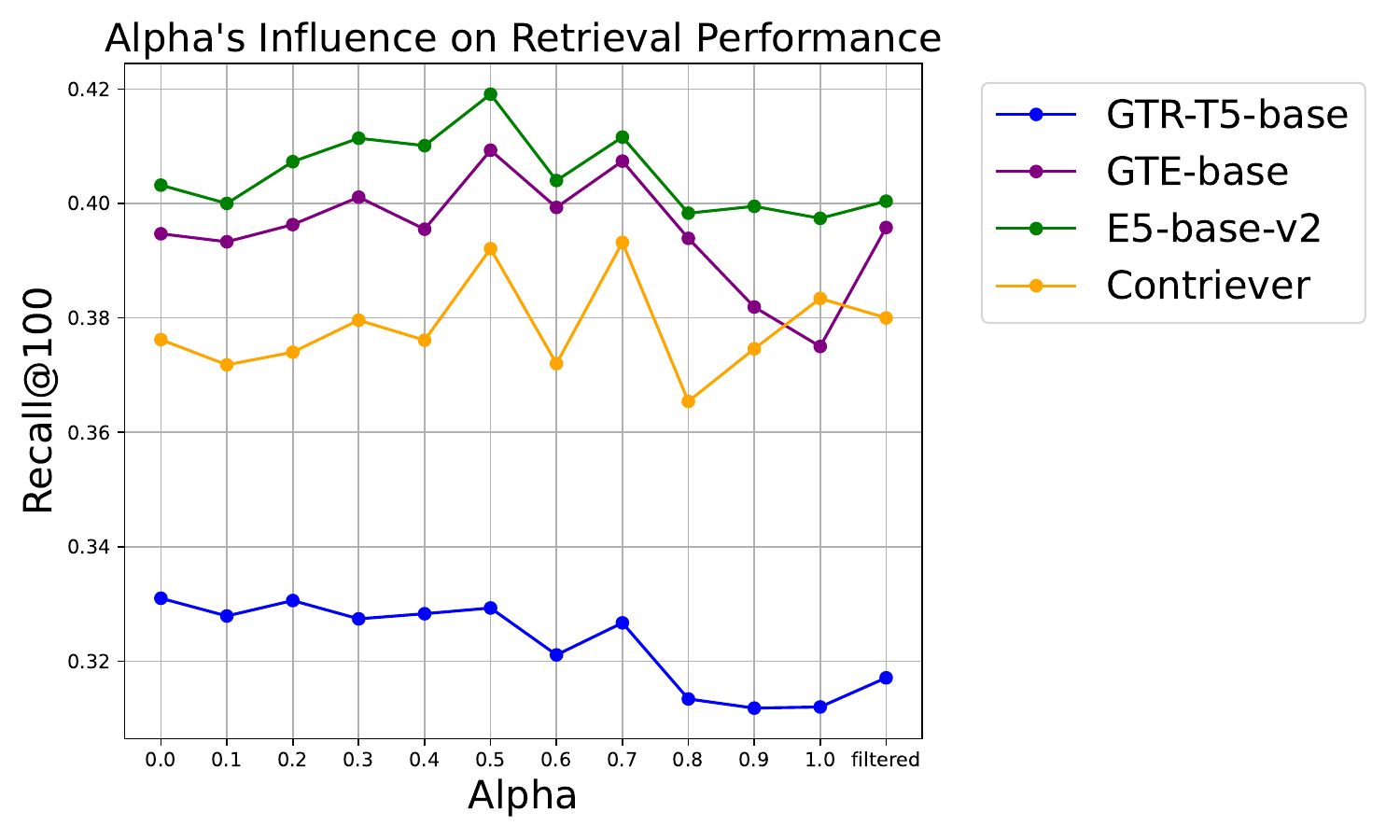}
\vspace{-0.5em}
  \caption{Retrieval Performance (Recall@100) of \ours with different $\alpha$ values on \questvar. Performance of \textit{SupCon} baseline is also put at the end. }
  \label{fig:hyperparam_study}
\vspace{-0.5em}
\end{figure}

To understand how including relevant queries in the same mini-batch affects the performance of the in-batch contrastive learning process, we study the effect of the hyperparameter \( \alpha \), which controls the proportion of random samples versus groups of related queries. We analyze its effect on both model performance and the coherence of embeddings. 
To measure the coherence of embeddings, within each group, we compute the pairwise similarity between queries that share the same atomic sub-queries base but differ in logical connectives. Formally, for a group \( G = \{q_1, q_2, \dots, q_n\} \) of queries sharing the same atomic sub-queries, we define the average embedding similarity as:

{\small
\begin{equation}
\text{AvgGroupSim}(G) = \frac{2}{|G|(|G|-1)} \sum_{i < j} \text{sim}(f(q_i), f(q_j))
\end{equation}
}

where \( f(q) \) denotes the embedding of query \( q \), and \( \text{sim}(\cdot, \cdot) \) is the cosine similarity between embeddings. 

As shown in Figure~4, the performance experiences a slight drop when \( \alpha \) increases from 0.8 to 1.0. This phase corresponds to augmenting the training data while still relying on the random batching strategy. However, as \( \alpha \) decreases from 0.7 to 0.5—indicating a greater use of related queries groups—the performance significantly improves, outperforming the baseline. Beyond this point, as \( \alpha \) continues to decrease, we observe a mild decline in performance.

These results suggest that while sampling logically related query groups enhances the model's ability to distinguish queries with different logical structures, relying solely on this sampling method may hurt its ability to generalize over semantic similarity. 

\section{Related Work}

\subsection{Dense Retrieval}

Dense Passage Retriever (DPR) \citep{karpukhin2020dense} pioneered the approach of dense retrieval by encoding queries and passages into dense vector representations. Subsequently, Generalizable T5 Retriever (GTR) \citep{ni2021large} uses a large pre-trained language model (T5) to encode texts, thus capturing richer semantic contexts. Contriever \citep{izacard2021unsupervised} and E5 \citep{li2023towards} introduces an unsupervised and weakly supervised framework that trains on a large corpus. General Text Embedding (GTE) \citep{wang2022text} adapts additional data sources on both pre-training and fine-tuning. SimCSE \citep{gao2021simcse} introduces a framework that leverages data augmentation techniques to generate effective sentence embeddings through contrastive learning without additional labeled data. However, these methods focus solely on surface-level semantic similarity and fail to account for the underlying logical connectives present in complex queries.

\subsection{Retrieval on Logical Queries}

To examine the performance of these dense retrievers on logical queries, \citep{malaviya2023quest} creates the \quest dataset, which consists of 3357 queries that map to a set of entities corresponding to Wikipedia documents. The BoolQuestions \citet{zhang2024boolquestions} dataset also contains logical queries and is constructed upon MSMarco \citep{nguyen2016ms}. There have also been attempts to improve retrieval performance on queries with implicit logical operators. \citet{mai2024setbert} uses GPT-3.5 to generate over 150,000 additional training samples to fine-tune a BERT model \cite{devlin2019bert} with a customized contrastive loss. \citet{krasakis2025constructing} interpretable representations through Learned Sparse Retrieval (LSR). In contrast, our approach directly encodes logical structure through contrastive learning with logic-based groups and constraint-based objectives, without relying on expensive synthetic data or lexical heuristics. 

\subsection{Contrastive learning}

Contrastive learning's efficacy relies heavily on the selection of negative samples, and several approaches \citet{robinson2020contrastive, zhang2025chain, xiong2020approximate} select and update hard negatives based on document embeddings distance or BM25 \cite{robertson2009probabilistic} during training. To avoid the huge costs in continuously selecting hard negatives during training time, \citep{hofstatter2021efficiently, sachidananda2023global, ma2024mode, solatorio2024gistembed, morris2024contextual} have explored optimizing batch strategies beforehand. Most recently, \citet{morris2024contextual} uses K-Means clustering on text embeddings to find the hardest possible batch to train the model. In contrast, the sampling strategy of our work is based on the logical structure of the queries instead of semantic similarity, and our mix strategy also enables the model to learn both semantics and logical distinctions. 

\citet{roth2004linear, roth2007global} introduce the early frameworks for joint constrained learning by formulating entity and relation identification as a global inference problem. This paradigm has also been applied to the training of neural networks by \citep{li2019logic}. This paradigm is further applied to information extraction tasks, such as temporal relation extraction \citet{wang2020joint} and Emotion-Cause Pair Extraction \citet{feng2023joint}. We extend the usability of this framework from information extraction tasks to the text retrieval task by utilizing the subset and the exclusion relations between in-batch queries. 
\section{Conclusions}

We present \ours, a logically-informed contrastive learning framework designed to enhance the performance of dense retrievers on queries with logical connectives. \ours introduces a novel group batching strategy that enables the model to learn distinguishable representations for queries that are logically different yet semantically similar. In addition, we leverage the logical relationships among in-batch queries through a joint constrained learning objective, ensuring logical consistency in representation learning. Our experiments on the task of retrieving Wikipedia entities given complex logical queries demonstrate the effectiveness of \ours over standard supervised contrastive learning. We further provide detailed analysis showing that \ours significantly improves retrieval performance on queries involving negation. Finally, a hyperparameter study illustrates how tuning batching randomness influences the learning of semantic and logical information.

\section*{Limitations}
Our work has the following limitations. First, due to limited computational resources, our experiments are conducted using moderate-sized bi-encoder models with approximately 110M parameters and an effective batch size of 32. Second, as existing benchmarks primarily focus on English, the cross-lingual generalizability of \ours remains an open question.  


\bibliography{custom, anthology}

\newpage
\appendix
\section{Appendix}

\subsection{Details On \questvar}
\label{app:questvar-details}

Since not all the companion queries exist in the dataset, we develop the following automated pipeline to augment the original training set. 

To mitigate this issue, we derived the missing base atomic queries from Wikipedia category names, which are hand-curated natural language labels assigned to groups of related documents in Wikipedia. We strictly follow the data construction procedure of Quest so that our extracted atomic queries match the original complex queries' decompositions.

After obtaining atomic queries, we then proceed to construct related complex queries from them. We limit our related complex queries to one of the 6 templates used in the original Quest dataset, as shown in Table \ref{tab:data_statistics}. Then, for each new complex query $A \circ_1 B\circ_1 C$, we derived its ground truth document set from the ground truth document sets of its atomic queries $A$, $B$, and $C$, by applying the same operations on its answer document sets.  

When combining these sub-queries and selected operators into a natural language, we first combine sub-queries through fixed templates, such as \textit{\{\} and \{\}, but not \{\}}. Then, to ensure that the newly constructed queries closely resemble real-world queries, we also prompted an LLM, to paraphrase the templatically generated query to fluent natural language. 

Besides augmenting the training dataset, we also augment the validation and testing datasets to provide a more comprehensive evaluation following the same rules.



To organize training batches, we cluster queries that share the same set of sub-queries \( \{A, B, C\} \). Specifically, we identify two types of queries to include in the cluster. The first type consists of other queries that involve the same sub-query components \( \{A, B, C\} \) but are connected with different logical operators; this encourages the model to differentiate between queries that are semantically similar but logically distinct. To avoid overpopulating our batch, we only add new complex queries that belong to one of Quest's original templates, which are $A \cap B \cap C$, $A \cap B \setminus C$, and $A \cup B \cup C$. The second type consists of the individual sub-queries \( A \), \( B \), and \( C \); this promotes the model's ability to understand the relationship between a complex query and its components, recognizing how sub-queries and logical operators combine to form the overall query meaning.

For example, given the query in the previous example, \textit{Orchids of Indonesia}, we include

\begin{itemize}[leftmargin=*,nosep]
    \item Other queries involving the same sub-queries but different logical structures
    \begin{itemize}
        \item \textit{Orchids of Indonesia and Malaysia and Thailand} ($A \cap B \cap C$)
        \item \textit{Orchids of Indonesia or Malaysia or Thailand} ($A \cup B \cup C$)
    \end{itemize}
    \item The individual sub-queries themselves:
    \begin{itemize}
        \item \textit{Orchids of Indonesia} (\( A \))
        \item \textit{Orchids of Malaysia} (\( B \))
        \item \textit{Orchids of Thailand} (\( C \))
    \end{itemize}
\end{itemize}



\end{document}